\documentclass[prb,twocolumn,aps,showpacs,fixfloats]{revtex4}
\usepackage{graphicx}
\usepackage{bm}
\usepackage{amsmath,amssymb}
\usepackage{subfigure}
\usepackage{float}
\usepackage{latexsym}
\usepackage{color}
\usepackage{enumerate}
\usepackage{pdfpages}
\usepackage{tikz}
\usepackage{hyperref}
\usepackage{graphicx}
\usepackage{bm}
\usepackage{amssymb} 
\usepackage{amsmath}
\usepackage{subfigure}

\usepackage{relsize}

\begin{document}

\title{Landau-Zener transition with energy-dependent decay rate of the excited state }

\author{M. E. Raikh}
\affiliation{Department of Physics and
Astronomy, University of Utah, Salt Lake City, UT 84112}

\begin{abstract}
A remarkable feature of the Landau-Zener transition is insensitivity
of the survival probability to the decay rate, $\tau^{-1}$, of the excited state. Namely, the probability for a particle which is initially (at $t\rightarrow -\infty$) in the ground state to remain at $t\rightarrow \infty$ in the same state  is insensitive to $\tau^{-1}$ which is due to e.g. coupling to continuum  [V. M. Akulin and W. P. Schleich,  Phys. Rev. A {\bf 46}, 4110 (1992)].   
This insensitivity was demonstrated for the case when the density of states in the continuum is energy-independent.  We study the opposite  limit when the density of states in the continuum is a step-like function of energy.  As a result of this step-like behavior of the density of states,  the decay rate of a driven excited level experiences a jump {\em as a function of time} at certain moment $t_0$. We take advantage of the fact that the analytical solution at $t<t_0$  and at $t>t_0$ is known.  We show that the decay enters the survival  probability when $t_0$ is comparable to the transition time.

\end{abstract}

\maketitle

\section{Introduction}

According to the celebrated papers\cite{Landau,Zener,Majorana,Stukelberg} by Landau, Zener, Majorana, and Stuekelberg, as two levels are swept by each other with velocity, $v$, a particle remains in the level, which it occupied before the crossing, with probability $\exp\left(-\frac{2\pi J^2}{v}  \right)$, where $J$ is the tunneling amplitude between the levels at the point of crossing.  

\begin{figure}
	\label{F1}
	\includegraphics[scale=0.5]{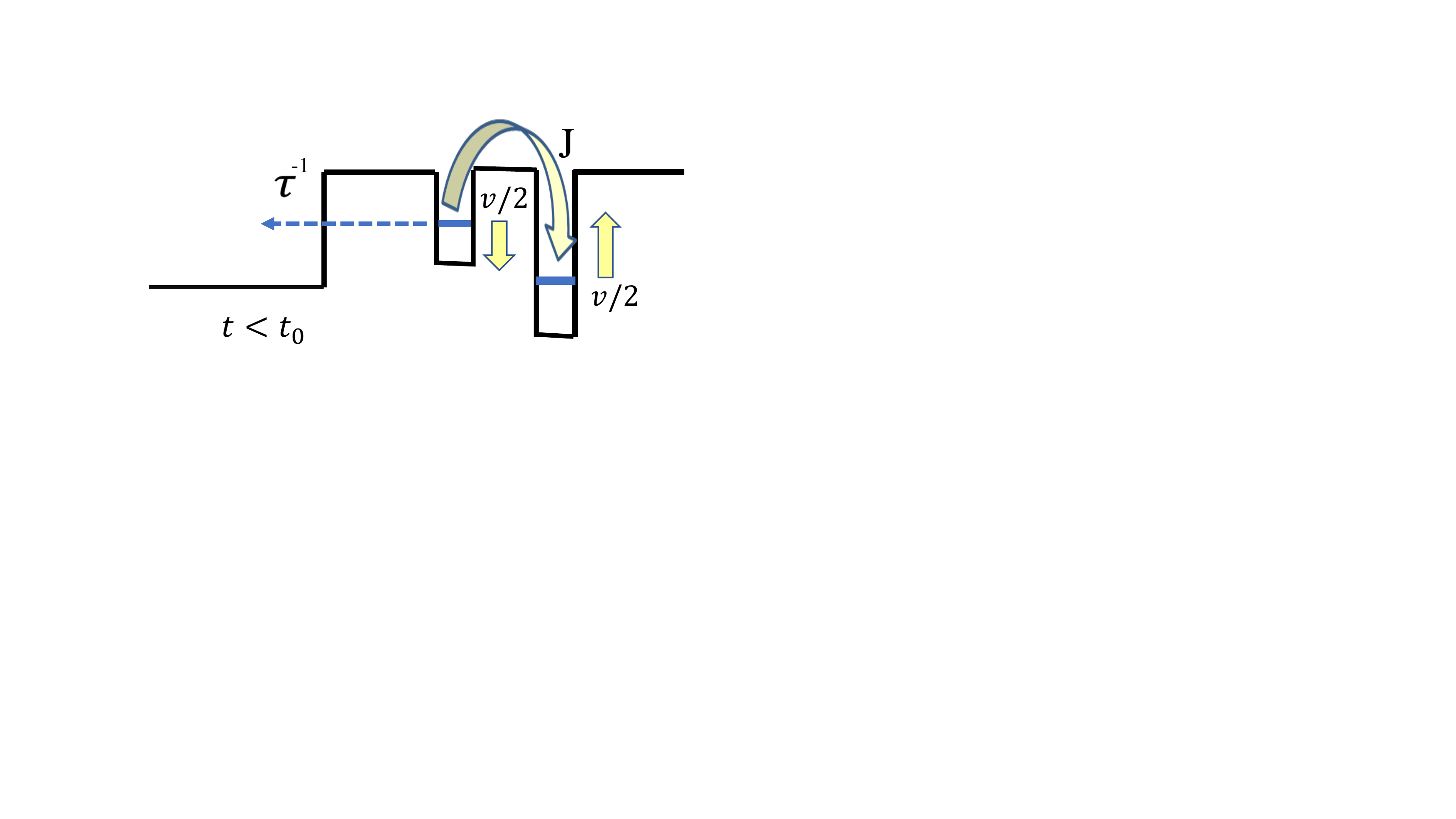}
	\caption{(Color online) Schematic illustration of the Landau-Zener transition with decaying excited state. Two levels are swept past each other with relative velocity $v$. Depending on the position of the left level, it can decay into continuum. At time $t<t_0$ the left level does not decay at all. At $t>t_0$ the electron in the left level can tunnel out into the continuum. The tunneling time is $\tau$. The question of interest is the dependence of the survival probability on $t_0 $. In the limit $t_0 \rightarrow -\infty$, the electron in the left level can tunnel out at all times. Then, as it was first demonstrated in Ref. \onlinecite{Akulin1992}, the tunneling time drops out from the survival probability.   }
\end{figure}

In the contemporary research the crossing of the driven levels is implemented in qubits based on superconducting circuits and quantum dot 
interferometers, see e.g. Refs. \onlinecite{review1}, \onlinecite{review2} for review.

It is remarkable that the Landau-Zener formula is robust with respect to the possibility of tunneling of a particle out of the excited state. As it was first pointed out in Ref. \onlinecite{Akulin1992}, see also Ref. \onlinecite {Vitanov1997}, for {\em any} $J$ the Landau-Zener formula remains applicable even when the tunneling time, $\tau$, is much shorter
than $\frac{J}{v}$, which is the characteristic time of the transition.
In the latter limit, each time an electron enters the excited state,
it typically  does not return to the initial state, but rather directly proceeds to the continuum. While the portion of events when the electron still returns is exponentially small, the Landau-Zener formula still holds despite the interference oscillations in the occupation of the excited state are completely washed
out.\cite{Vitanov1997}

While the result\cite{Akulin1992,Vitanov1997} seems counter-intuitive, it can be interpreted as a manifestation of the profound property of multistate Landau-Zener transition
established in Refs.~\onlinecite{Sinitsyn2004},~ \onlinecite{Shytov}. 
The key to this interpretation lies in the observation that the tunnel decay of the excited state can be modeled by the broadening of the level, corresponding to the excited state, into a multiplet of the closely spaced discrete levels.  Then the scenario of crossing of two levels  translates into the crossing of the driven ground-state level by the numerous excited-state levels. For this arrangement, it was demonstrated in Refs. \onlinecite{Sinitsyn2004},  \onlinecite{Shytov} that the survival probability is equal to the product of partial probabilities regardless of how close the levels forming the multiplet are spaced. With matrix element,  $J$,  being distributed between the sublevels, the width of the multiplet drops out of the product.

The fact that survival probability is independent of the tunneling width of the excited state poses a fundamental question:  will the decay enter the survival probability if it switches on  only above a certain energy, $\varepsilon_0$?  Another way to pose this question is: since the excited level is driven, the decay of the excited state switches on at certain time moment, $t_0$.  Will then $t_0$ enter the survival probability? This question is the focus of the present paper.  To address it,  we take advantage of the fact that 
the analytical solutions at $t<t_0$  and at $t>t_0$ are known.  The prime question is, certainly,  whether switching on the decay enhances or suppresses the survival probability. Note also, that here is  a natural scale for $t_0$, which is  the characteristic time of the transition,  $\frac{J}{v}$.

\section{The Model}

It is convenient to reason within a concrete model of 
levels which are swept passed each other 
with a relative velocity, $v$. This model is illustrated in Fig. 1.  Assume that the right level evolves with time as

\begin{equation}
\label{plus}
\varepsilon_r(t)=\frac{vt}{2}
\end{equation}
and does not decay. The left  level evolves with time as
\begin{equation}
\label{minus}
\varepsilon_l(t)=-\frac{vt}{2}.
\end{equation}

As illustrated in Fig. 1, 
at time $t>t_0$ an electron can tunnel out of the left level  into the continuum, while at  $t<t_0$  the tunneling channel is shut off.  

Initially,  at $t\rightarrow -\infty$ an electron is located in the right level. According to  Refs. \onlinecite{Akulin1992} and \onlinecite{Vitanov1997}, at large positive $t_0 \gg \frac{J}{v}$, the tunnel decay drops out of the survival probability, $\exp\left(-\frac{2\pi J^2}{v}  \right)$.

Equally,  at  large negative  $t_0\ll -\frac{J}{v}$, the Landau-Zener transition  proceeds as if there was no decay. Thus, if we view the survival probability as a function of $t_0$, we expect that
the decay will enter this function only if $t_0$ falls into the interval $\sim \frac{J}{v}$.

At all times, evolution  of the amplitude to find the electron in the right level is described by the equation

\begin{equation}
\label{right}
i\dot{a}=\frac{vt}{2}a+ Jb.
\end{equation}
 Concerning the amplitude to find electron 
 in the left level, before and after switching  evolves as
 \begin{align}
 \label{left}
 &i\dot{b}= -\frac{vt}{2}b+Ja,  \hspace{3mm}  t<t_0 \nonumber\\
 &i\left(\dot{b}+\frac{b}{\tau}\right) = -\frac{vt}{2}b+ Ja,~~t>t_0. 
\end{align}
The system Eqs. \ref{right} and \ref{left} should be solved under the condition $b(-\infty)=0$,  i.e. under the condition that the electron is initially in  the right level. The   solution at $t<t_0$ has the textbook form

\begin{equation}
\label{form}
a(t)=D_{\nu}(z),~~   b(t)=-i\sqrt{\nu}D_{\nu-1}(z),
\end{equation}
where the argument of the parabolic
cylinder function, $D_{\nu}$, is defined as
\begin{equation}
\label{z}
z=t\sqrt{v}e^{\pi i/4},
\end{equation}
 while the parameter $\nu$ is  expressed via $J$ and $v$ as
 \begin{equation}
 \nu=-\frac{iJ^2}{v}.
 \end{equation}
 To find the solution at $t>t_0$ we make the following substitution in Eqs. \ref{right},  \ref{left}
 \begin{equation}
 \label{substitution}
 a(t)=a_0(t)\exp\left(-\frac{t}{2\tau}\right),~~b(t)=b_0(t)\exp\left(-\frac{t}{2\tau}\right).
 \end{equation}
 Upon this substitution, we arrive to the following  system of equations for $a_0$(t) and $b_0(t)$ 
 \begin{align}
 \label{system}
 &i\dot{a}_0=\left(\frac{i}{2\tau} +\frac{vt}{2}   \right)a_0 +Jb_0, \nonumber\\
 &i\dot{b}_0=-\left(\frac{i}{2\tau} +\frac{vt}{2}   \right)b_0+Ja_0.
 \end{align}
 We see that the system Eq. \ref{system}  reproduces the system for $a(t)$ and $b(t)$ upon the replacement $t\rightarrow t+\frac{i}{v\tau}$.
 
 General solution of the system Eq.  \ref{system} is given by the linear combination of the parabolic  cylinder functions of the shifted argument
 
 \begin{align}
\label{SHIFTED}
 &a_0(t)=AD_{\nu}\left[\left(t+\frac{i}{v\tau}\right)\sqrt{v}e^{\pi i/4}\right]\nonumber\\
& +BD_{\nu}\left[-\left(t+\frac{i}{v\tau}\right)\sqrt{v}e^{\pi i/4}\right],
 \end{align}
 
\begin{align}
\label{SHIFTED1}
 &b_0(t)=-i\sqrt{\nu}\Biggl\{
 AD_{\nu-1}\left[\left(t+\frac{i}{v\tau}\right)\sqrt{v}e^{\pi i/4}\right]\nonumber\\
& -BD_{\nu-1}\left[-\left(t+\frac{i}{v\tau}\right)\sqrt{v}e^{\pi i/4}\right]
\Biggr\}.
 \end{align}
 Constants $A$ and $B$ are determined from the continuity of $a(t)$ and $b(t)$ at $t=t_0$. The corresponding system reads
 
 \begin{align}
 \label{AandB}
 &D_{\nu}(z_0)e^{\frac{t_0}{2\tau}}=AD_{\nu}(z_0')+BD_{\nu}(-z_0'),
 \nonumber\\
 &D_{\nu-1}(z_0)e^{\frac{t_0}{2\tau}}=AD_{\nu-1}(z_0')-BD_{\nu-1}(-z_0').
 \end{align}
where the arguments $z_0$ and $z_0'$ are defined as
\begin{equation}
\label{arguments}
z_0=t_0\sqrt{v}e^{\pi i/4},
~~z_0'=z_0+\frac{i}{\sqrt{v}\tau}e^{\pi i/4}.
\end{equation}Solving the system Eq. (\ref{AandB}) yields the following expressions  for the constants $A$ and $B$ 
 \begin{align}
  &A=e^{\frac{t_0}{2\tau}}
 \frac{D_{\nu}(-z_0')D_{\nu-1}(z_0)+D_{\nu }(z_0)D_{\nu-1}(-z_0')}{D_{\nu}(-z_0')D_{\nu-1}(z_0')+D_{\nu}(z_0')D_{\nu-1}(-z_0') },  \label{A+B}
\\
&B=-e^{\frac{t_0}{2\tau}}
 \frac{D_{\nu-1}(z_0)D_{\nu}(z_0')-D_{\nu }(z_0)D_{\nu-1}(z_0')}{D_{\nu}(-z_0')D_{\nu-1}(z_0')+D_{\nu}(z_0')D_{\nu-1}
(-z_0') } \label{A1+B1}.
 \end{align}
 In the absence of decay, $\tau \rightarrow \infty$,   the dependence on the moment, $t_0$,  should drop out.  Indeed, setting $z_0=z_0'$ in  Eqs. \ref{A+B}, 
 \ref{A1+B1}, we find $A=1$, $B= 0$.

The quantity we are interested in is the survival probability, which we define as the ratio of probabilities to find the particle  in  the left  level at $z\rightarrow \infty$ and $z\rightarrow -\infty$.  With the help of the explicit form of $a(t)$ given by Eq. (\ref{SHIFTED}), we find

\begin{equation}
\label{Q}
Q_{LZ}-\frac{\Big{\vert}A\left(e^{-\frac{t}{2\tau}} D_{\nu} (z)\right)_{t\rightarrow \infty}+B\left( e^{-\frac{t}{2\tau}} D_{\nu} (-z)     \right)_{t\rightarrow \infty}\Big{\vert}^2}{\Big{\vert}D_{\nu}(z)\Big{\vert}_{t\rightarrow -\infty}^2    }.
\end{equation}
Using the large-argument asymtotes of the parabolic-cylinder functions\cite{Bateman},  we obtain\cite{Zhuxi}
 
  \begin{equation}
  \label{general}
  Q_{LZ}=\vert A   \vert^2 e^{-2\pi|\nu|}
  +\left( A^{\ast}B+B^{\ast}A\right)e^{-\pi|\nu|} +\vert B\vert^2.
  \end{equation}
 Equations Eqs. (\ref {A+B}), (\ref{A1+B1}) and (\ref{general}) yields a formal solution to the problem. We will analyze it in the most interesting limit $|t_0|  \ll \frac{J}{v}$,  when the decay switches on  in the course  of the transition. Then, as demonstrated in the Appendix, the expressions for the amplitudes $a(t)$ and $b(t)$ before switching of the decay can be simplified as 
 
 \begin{align}
 \label{LAMBDA}
 &a(t)=\lambda e^{iJt}+\lambda^{-1}e^{-iJt}, \nonumber\\
 &b(t)=-\lambda e^{iJt} +\lambda^{-1}e^{-iJt}.
 \end{align}
  
 Here  $\lambda=\exp\left( -\frac{\pi J^2}{2v }\right)
 =\exp\left(-\frac{\pi \nu}{2}\right)$.
 
 It is convenient to cast $a(t_0^{-})$ and $b(t_0^{-})$, which are the amplitudes
 at $t=t_0^{-}$, 
 in a more concise form by introducing a notation
 
 \begin{equation}
\label{kappa}
\kappa_0=\lambda e^{iJt_0}.
\end{equation}
 Then Eq. (\ref{LAMBDA}) takes the form
 
 \begin{align}
 \label{LAMBDA1}
 &a(t_0^{-})=\kappa_0+\frac{1}{\kappa_0}, \nonumber\\
 &b(t_0^{-})=-\kappa_0+\frac{1}{\kappa_0}.
 \end{align}
 
Similarly, upon   switching on the tunneling, the simplified  expressions for $a(t)$ and $b(t)$ read
  \begin{align}
 \label{more}
 &a(t)=e^{-\frac{t}{2\tau}}
 \Bigg[A\Big\{\lambda  e^{iJ\left(t+\frac{i}{v\tau}\right)}  +\lambda^{-1}e^{-iJ\left(t+\frac{i}{v\tau}\right)} \Big\}\nonumber\\
 &+B\Big\{\lambda  e^{-iJ\left(t+\frac{i}{v\tau}\right)}  +\lambda^{-1}e^{iJ\left(t+\frac{i}{v\tau}\right)} \Big\}\Bigg], \nonumber\\
 &b(t)=e^{-\frac{t}{2\tau}}
  \Bigg[-A\Big\{\lambda  e^{iJ\left(t+\frac{i}{v\tau}\right)}  -\lambda^{-1}e^{-iJ\left(t+\frac{i}{v\tau}\right)} \Big\}\nonumber\\
 &+B\Big\{\lambda  e^{-iJ\left(t+\frac{i}{v\tau}\right)}  -\lambda^{-1}e^{iJ\left(t+\frac{i}{v\tau}\right)} \Big\}\Bigg].
 \end{align}
 
 Again, the above expressions assume a concise form with the help of the auxiliary notations
 
 \begin{align}
 \label{notation}
 &\kappa=\lambda \exp\left\{ iJ\Big(t_0+\frac{i}{v\tau}\Big)     \right\}, \nonumber\\
 & \mu=\lambda \exp\left\{  - iJ\Big(t_0+\frac{i}{v\tau}\Big)     \right\}.
\end{align}

With the help of these notations, the amplitudes $a(t)$ and $b(t)$ at
$t=t_0^{+}$ can be written as 
\begin{align}
\label{help}
&a(t_0^{+})=e^{-\frac{t_0}{2\tau}} \left[Af_1+Bg_1 \right],     \nonumber\\
&b(t_0^{+})=e^{-\frac{t_0}{2\tau}}\left[Af_2+Bg_2       \right],
\end{align}
where the coefficients $f_1$, $f_2$, $g_1$, and $g_2$ 
are given by

\begin{align}
\label{cefficients}
&f_1=\kappa+\frac{1}{\kappa}, \hspace{3mm} f_2=-\kappa+\frac{1}{\kappa}, 
\nonumber\\
&g_1=\mu+\frac{1}{\mu}, \hspace{3mm} g_2=\mu-\frac{1}{\mu}.
\end{align}
From the continuity conditions $a(t_0^-)=a(t_0^+)$,   $b(t_0^-)=b(t_0^+)$
we infer the following expressions for $A$ and $B$
\begin{align}
\label{formal}
&A=e^{\frac{t_0}{2\tau}} \frac{g_1 b(t_0^-)- g_2b(t_0^{-})        }
{f_2g_1-f_1g_2},
\nonumber\\
&B=e^{\frac{t_0}{2\tau}}\frac{f_2a(t_0^-)-f_1b(t_0^{-})        }{f_2g_1-f_1g_2}.
\end{align}
Note that the combination in denominators in Eq. (\ref{formal}) is equal to
\begin{equation}
\label{denominators}
f_2g_1-f_1g_2=2\mu\left(\frac{1}{\kappa}-\kappa     \right).
\end{equation}

Then the system  Eq. (\ref{AandB}) yields the  following results for the coefficients $A$ and $B$ 

 \begin{equation}
 \label{AFINAL}
A=
\frac{\lambda^4 e^{\frac{2J}{v\tau}}-1}{\lambda^4-1}
\exp\Big[\frac{t_0}{2\tau}-\frac{J}{v\tau}\Big].
\end{equation}
 
 \begin{equation}
 \label{Bfinal}
 B=-2\lambda^2\exp{\left(\frac{t_0}{2\tau}\right)}
 \frac{\sinh \left(\frac{J}{v\tau}\right)}
{\lambda^4-1}.
 \end{equation}

Substituting these values into Eq. (\ref{Q}), we arrive at the final result
\begin{equation}
\label{combination}
Q_{LZ}=\Big[\lambda^2A+B\Big]^2=e^{-2\pi|\nu|}
\exp\left[   \frac{2}{\tau}
\left( \frac{J}{v}+\frac{t_0}{2}  \right)\right].
\end{equation}
While $e^{-2\pi|\nu|}$ is a conventional Landau-Zener result,
the second exponent describes the effect of the tunnel decay
on the survival probability and on the moment, $t_0$, of the 
switching on the decay.

\section{Concluding Remarks}

An interesting feature of the result obtained is that there is a special 
time moment $t_0= - \frac{2J}{v}$ of switching  on the decay. For this $t_0$ the survival probability does not decay on the tunnel decay at all. Qualitatively, the meaning of the ratio ${J}{v}$ is the characteristic time of the Landau-Zener transition.
As seen from Eqs. (\ref{right}) and (\ref{left}),  the ratio $\frac{J}{v}$ is the time 
when the amplitudes $a(t)$ and $b(t)$ are of same order.
Concerning the dependence of $Q_{LZ}$ on the tunneling   
rate $\tau^{-1}$ ,  it is seen from Eq.  \ref{combination}     the
characteristic $\tau$ is also $\frac{J}{v}$.  Note that, see Eq. \ref{LAMBDA},   the probabilities  $|a(t)|^2$ and 
$|b (t)|^2$ contain a constant part 
$\frac{1}{\lambda^2} +\lambda^2$ plus  oscillatory parts $\pm 2\cos Jt$.  When the survival probability is small,  $\lambda \ll 1$, oscillatory parts constitute a small correction. At  $\tau \lesssim \frac{J}{v}$ the oscillations are completely washed out by the decay.\cite{Vitanov1997}
If $\tau$ is constant, this washout does not  affect $Q_{LZ}$.
On the contrary,   the second factor in Eq.  (\ref{combination})  is entirely due to the      .abrupt switching on of the decay. 

It is instructive to compare our results with 
Ref. \onlinecite{Rajesh} where the decay also 
enters into the survival probability. In 
Ref. \onlinecite{Rajesh} the left and right 
levels formed doublets.
The states of the right doublet, emulating the
initial state of the Landau-Zener transition,
did not decay, while both states of the
right doublet, emulating the final state, 
did decay into continuum.
It was shown in Ref. \onlinecite{Rajesh},
that, unlike the conventional transition,
the decay rates of the final states entered 
the survival probability, if they are different. 
By contrast, in the present manuscript we consider
a conventional Landau-Zener transition, but allow 
{\em both} mutually crossing levels to decay.
Calculation in  Ref. \onlinecite{Rajesh} illustrates
the difference between the integrable and non-integrable
Landau-Zener models: in integrable models the tunnel decay
drops out of the survival probability, while in non-integrable
models the decay enters the survival probability,  $Q_{LZ}$, {\em explicitly}. In this regard, the message of the present paper can be formulated as follows: even  an integrable model,  with a tunnel rate being a function of energy,  it enters  explicitly into $Q_{LZ}$.

\vspace{5mm}

{\bf Appendix: derivation of Eq.18 }

\vspace{5mm}

We start from the integral representation\cite{Bateman} of the parabolic cylinder function 

\begin{equation}
\label{representation}
D_{\nu}(z)=\left(\frac{2}{\pi}   \right)^{1/2}e^{\frac{z^2}{4}}\int\limits_0^{\infty}du e^{-\frac{u^2}{2}}u^{\nu}
\cos\left (zu-\frac{\pi u}{2}\right).
\end{equation}

Following Ref.  [\onlinecite{Zhuxi}], it is convenient to divide
the integral Eq. (\ref {representation}) into two contributions
\begin{equation}
	\label{two}
	D_{\nu}(z)=	I_{+}(z)e^{\frac{i\pi\nu}{2}} +
	I_{-}(z)e^{-\frac{i\pi\nu}{2}},
\end{equation}
where the functions $I_+(z)$ and $I_-(z)$ are defined as
\begin{equation}
\label{definition}
I_\pm(z)=\Bigl(\frac{1}{2\pi} \Bigr)^{1/2}e^{\frac{z^2}{4}}\int\limits_0^{\infty}	du~
u^{\nu}e^{-\frac{u^2}{2}\pm iuz}.
\end{equation} 
Both integrals are evaluated with the help
of steepest descent approach. The corresponding saddle points are given by
\begin{equation}	
	\label{saddle}
	u_{\pm}=\frac{iz}{2} \pm \frac{(4\nu-z^2)^{1/2}}{2}.
\end{equation}
Performing the Gaussian integration around the saddle points, we arrive to the following asymptotic expressions for $I_+$,  $I_-$
\begin{align}
\label{I+}
&I_+(z)\approx (4\nu-z^2)^{-1/4} \exp\Biggl[\frac{1}{4}iz(4\nu-z^2)^{1/2}  -\frac{1}{2}\nu  \Biggr]\nonumber\\
&\times	\Biggl[\frac{1}{2}iz+\frac{1}{2}(4\nu-z^2)^{1/2} \Biggr]^{\nu+\frac{1}{2}},
\end{align}
	
\begin{align}
	\label{I-}
	&I_-(z)\approx (4\nu-z^2)^{-1/4} \exp\Biggl[-\frac{1}{4}iz(4\nu-z^2)^{1/2}  -\frac{1}{2}\nu  \Biggr]\nonumber\\
	&\times	\Biggl[-\frac{1}{2}iz+\frac{1}{2}(4\nu-z^2)^{1/2} \Biggr]^{\nu+\frac{1}{2}}.
\end{align}	
The condition of applicability of the steepest descent method is that the typical value of   $\left(u-u_+\right)$
contributing to the integral Eq.  (\ref{definition})
is much smaller than $u_+$. 
With characteristic time of the transition being
$\frac{J}{v}$, we conclude that characteristic $z$ is $\sim \frac{J}{v^{1/2}}\sim |\nu|^{1/2}$.  Thus, for $|\nu|\gg 1$, when the position of the saddle point is $u_+ \sim |\nu|^{1/2} \gg 1$ the saddle-point result applies not only for large, but
for arbitrary $z$.\cite{Mishchenko}

To derive Eq. (\ref{LAMBDA}), we substitute Eqs. (\ref{I+}), (\ref{I-}) into 
Eq. (\ref{two}). This yields
\begin{align}
\label{INTERMS}	
&D_{\nu}(z)=e^{\frac{\pi i}{8}+\frac{iJ^2}{2v}}\biggl(\frac{v}
{W^2(t)}   \biggr)^{1/4}
\Biggl(\frac{e^{-\frac{\pi i}{4}}}{v^{1/2}}\Biggr)^{\frac{1}{2}-
	\frac{iJ^2}{v}}
\nonumber\\
&\times \Bigg\{e^{{\frac{it}{4}}W(t)}\Bigg(\frac{-vt+
	W(t)}{2}
\Bigg)^{\frac{1}{2}-\frac{iJ^2}{v}}
e^{-\frac{\pi J^2}{2v}}
\nonumber\\
&+e^{{-\frac{it}{4}}W(t)}\Bigg(\frac{vt+
	W(t)}{2}
\Bigg)^{\frac{1}{2}-\frac{iJ^2}{v}}
e^{\frac{\pi J^2}{2v}}    \Bigg\},	
\end{align}
where $W(t)=\left(4J^2+v^2t^2\right)^{1/2}$.   Taking the limit $vt \ll J$ and recalling that $\exp\left(-\frac{\pi J^2}{2v}\right)=\lambda$, we recover Eq. (\ref{LAMBDA}).

\end{document}